\documentclass[aps,prl,reprint,superscriptaddress,amsmath,amssymb,floatfix]{revtex4-2}
\usepackage{graphicx}
\usepackage{bm}
\usepackage{microtype}
\usepackage[hidelinks]{hyperref}
\hypersetup{pdftitle={Quantum Metric Scaling and Its Exceptions in Classical Flat Bands},
 pdfauthor={Kin Hung Fung; K. T. Law; C. T. Chan}}
\newcommand{\ellqm}{\ell_{\mathrm{QM}}}

\newcommand{\eh}{\widehat{\varepsilon}}
\newcommand{\wh}{\widehat{w}}
\newcommand{\eps}{\varepsilon}
\makeatletter
\newcommand{\doilink}[2]{\hyper@linkurl{#2}{https://doi.org/#1}}
\makeatother

\begin{document}

\title{Quantum Metric Scaling and Its Exceptions in Classical Flat Bands}

\author{Kin Hung Fung}
\email{funguiuc@gmail.com}
\affiliation{Department of Physics, The Hong Kong University of Science and Technology, Hong Kong, China}
\author{K. T. Law}
\affiliation{Department of Physics, The Hong Kong University of Science and Technology, Hong Kong, China}
\author{C. T. Chan}
\email{phchan@ust.hk}
\affiliation{Institute for Advanced Study, The Hong Kong University of Science and Technology, Hong Kong, China}

\begin{abstract}
The quantum metric length has been proposed as the localization length scale in one-dimensional flat bands. Here, we study this quantum metric scaling and the exceptions to it in a classical flat band formed by a locally resonant mass--spring system. We find that the localization length follows the quantum metric length near the flat band but departs from this proportionality at the exact flat-band frequency. Selective mass disorder also produces divergent localization lengths or arbitrarily short tails near the flat-band frequency without changing the clean metric. These exceptions show that disorder can select localization scales beyond those predicted by the quantum metric alone.
\end{abstract}
\maketitle

Quantum geometry reveals a rich connection between the geometric structure of Bloch eigenstates and physical properties beyond the energy dispersion~\cite{Yu2025}. The quantum metric measures the infinitesimal distance between neighboring Bloch eigenstates in momentum space~\cite{Provost1980}, and its Brillouin-zone integral determines the gauge-invariant part of the quadratic Wannier spread~\cite{Marzari1997}, linking momentum-space geometry to real-space spreading. Flat bands provide a particularly revealing setting for studying how band geometry influences the spatial extent of states. Even when the group velocity vanishes, the Bloch eigenvectors can retain momentum dependence and a tunable finite quantum metric~\cite{Luo2025,LiuHotspots2026}. It has been shown that disorder can mix the macroscopically degenerate flat-band states and can generate unconventional localization behavior~\cite{Chalker2010,Bodyfelt2014,Flach2014,Leykam2017,Cadez2021}, involving a strong dependence on frequency detuning~\cite{Ge2017}. The quantum metric also contributes to flat-band superfluid weight~\cite{Peotta2015} and optical and nonlinear responses~\cite{Ahn2022,Wang2023}. Related geometric lengths appear in superconducting coherence and proximity effects~\cite{ChenLaw2024,Li2025}, boundary and bound states~\cite{MaBoundary2025,Zhao2026}, and Friedel oscillations~\cite{MaFriedel2026}.

\begin{figure}[!t]
\centering
\includegraphics[width=\columnwidth]{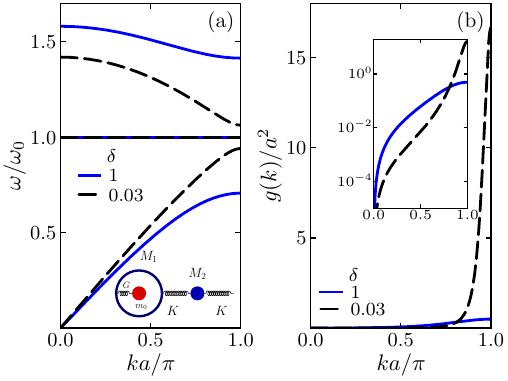}
\caption{\label{fig:model}Clean bands and quantum metric of the locally resonant mass--spring chain. (a) Bloch frequencies $\omega/\omega_0$ at $G/m_0=2K/M=\omega_0^2$, for gap-opening parameter $\delta=G/K=1$ (blue solid) and $0.03$ (black long-dashed). The middle band remains exactly flat at $\omega_0$, while decreasing $\delta$ narrows its separation from the dispersive bands. Inset: a unit cell of period $a$, with chain masses $M_1=M_2=M$ joined by springs $K$ and a resonator $m_0$ attached to $M_1$ by a spring $G$. (b) Flat-band quantum metric $g(k)/a^2$ from mass-normalized Bloch eigenvectors in the physical embedding $r_1=r_0=0$, $r_2=a/2$. Decreasing $\delta$ sharpens and enhances the peak near $k=\pi/a$, where $g(\pi/a)=a^2/(2\delta)$. The inset shows the same curves with a logarithmic vertical scale. Both panels use the same line styles and dimensionless wavevector $ka/\pi$.}
\end{figure}

Recent studies have shown that the localization length in disordered flat bands can be governed by the geometry of the Bloch states~\cite{Chau2026,Dai2026}. In this regime, known as \emph{quantum metric localization}, the localization length is proportional to the quantum metric length and remains nearly independent of disorder strength in medium to strong disorder regimes. For an isolated narrow band, increasing disorder beyond the bandwidth can halt the conventional decrease of the localization length and produce a plateau set by band geometry~\cite{Dai2026}. In real space, this behavior can be understood in terms of disorder-induced transitions between overlapping Wannier states, whose spatial spread is constrained by the quantum metric. The persistence of this geometric localization scale over a broad range of disorder strengths is termed \emph{quantum metric protection}~\cite{Dai2026}. This protection, which concerns a tunable spatial scale rather than a quantized response, can occur even in topologically trivial flat bands.

Classical wave systems offer an experimentally accessible and highly tunable setting for studying quantum metric localization. Flat bands have been realized in photonic lattices~\cite{Mukherjee2015,Vicencio2015}, phononic metamaterials~\cite{Samak2024}, and electrical circuits~\cite{ChaseMayoral2024}. Quantum-geometric quantities can be measured in wave experiments~\cite{LiuQiu2026}, and localization can be spatially resolved in photonic and elastic-wave systems~\cite{Schwartz2007,Hu2008}. Here, we study this quantum metric scaling and the exceptions to it in a classical flat
band formed by a locally resonant mass--spring system.

\textit{Classical flat-band model and quantum metric.---} We consider the locally resonant mass--spring chain in Fig.~\ref{fig:model}(a), with two masses $M_1=M_2=M$ per unit cell, connected by springs $K$, and a local resonator $m_0$ attached to each $M_1$ by a spring $G$~\cite{Liu2000,Fung2005}. It supports a tunable flat band while allowing mass disorder to be applied independently to each sublattice. With unit-cell period $a$ and Bloch displacement amplitudes $\bm U=(U_1,U_2,U_0)^T$, the Bloch equation is $D(k)\bm U=\omega^2\mathcal M\bm U$, where $\mathcal M=\operatorname{diag}(M,M,m_0)$ and
\begin{equation}
D(k)=\begin{pmatrix}
2K+G&-K(1+e^{-ika})&-G\\
-K(1+e^{ika})&2K&0\\
-G&0&G
\end{pmatrix}.
\label{eq:D}
\end{equation}
At $G/m_0=2K/M=\omega_0^2$, the middle band is exactly flat, with the flat-band eigenstate $\bm U_{\rm FB}(k)\propto(0,-\delta,1+e^{-ika})^T$. Here $\delta=G/K$ is the dimensionless gap-opening parameter, which controls the separation of the flat band from the dispersive bands. As shown in Fig.~\ref{fig:model}(a), increasing $\delta$ widens the band gaps while the middle band remains exactly flat.

To calculate the quantum metric, we normalize the states by the kinetic energy and rewrite the eigenvalue problem as a Hermitian problem $\mathcal M^{-1/2}D(k)\mathcal M^{-1/2}\bm V=\omega^2\bm V$, with $\bm V=\mathcal M^{1/2}\bm U$. The normalized flat-band eigenvector is proportional to $(0,-\sqrt{2\delta},2\cos(ka/2))^T$. Its quantum metric and the corresponding quantum metric length are~\cite{Marzari1997,Chau2026,Dai2026}
\begin{align}
g(k)&=\frac{2\delta a^2\sin^2(ka/2)}{[2\delta+4\cos^2(ka/2)]^2},\notag\\
\ellqm&=\frac{1}{2\pi}\int_{-\pi/a}^{\pi/a}g(k)\,dk
=\frac{a}{4\sqrt{\delta(2+\delta)}}.
\label{eq:metric}
\end{align}
The $k$ dependence of $g(k)$ is also illustrated in Fig.~\ref{fig:model}(b), showing a diverging value at the zone boundary when $\delta$ is small. The Brillouin-zone-averaged metric gives the gauge-invariant contribution to the quadratic Wannier spread. Dividing that spread by the period $a$ gives $\ellqm$. 

\begin{figure*}[!t]
\centering
\includegraphics[width=\textwidth]{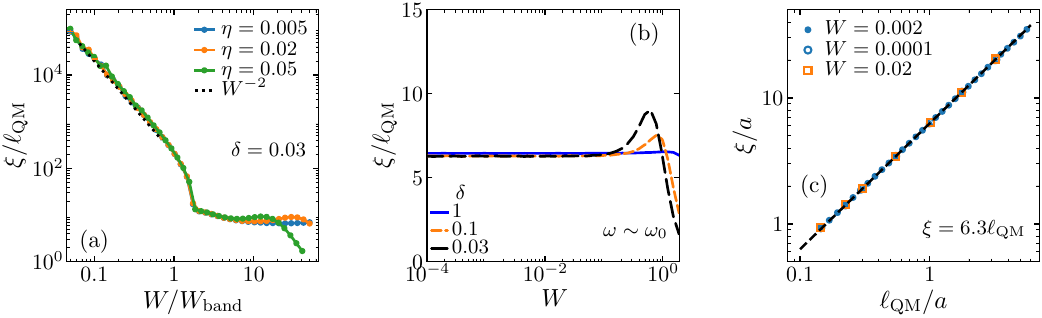}
\caption{\label{fig:disorder}Quantum metric localization near the flat band under independent uniform mass disorder. (a) Small residual dispersion at $\delta=0.03$: detuning $m_0\to m_0(1+\eta)$ produces bandwidth $W_{\rm band}=\Delta(\omega^2)/\omega_0^2$. The squared frequency is chosen at the residual-band center plus $0.2W\omega_0^2$. The conventional $W^{-2}$ regime (dotted guide) crosses over to a localization-length plateau on the scale of $\ellqm$ when disorder exceeds the residual bandwidth. The lengths are normalized by $\ellqm$ of the exactly flat parent band. Error bars at the six points with the smallest disorder strengths are obtained by propagating the standard error of the mean transfer rate over eight realizations. (b) For an exactly flat parent band, the localization length forms plateaus at $\omega\sim\omega_0$ ($\eh=0.2$) for $\delta=1$, $0.1$, and $0.03$. At $\delta=1$, the plateau extends to $W=1.9$. (c) Localization length as a function of the exact quantum metric length at $\omega\sim\omega_0$ ($\eh=0.2$). The data include 31 values of $\delta$ at $W=0.002$, supplemented by seven values of $\delta$ at $W=10^{-4}$ and $0.02$. The dashed guide $\xi=6.3\ellqm$ is specific to this detuning and disorder ensemble.}
\end{figure*}

\textit{Quantum metric localization near the flat band.---} To study localization, we introduce mass disorder by setting the individual masses (labeled by the sublattice index $\alpha$ and the unit-cell index $n$) to $m_{\alpha,n}=m_\alpha(1+W\wh_{\alpha,n})$, where the independent random variables $\wh$ are uniformly distributed on $[-1/2,1/2]$. This reference ensemble describes independent mass loading at fixed spring constants. 

To see how the flat-band behavior emerges from conventional localization, we first allow the middle band to acquire a small residual dispersion. Detuning the resonator mass, $m_0\to m_0(1+\eta)$, gives the middle band a width $W_{\rm band}=\Delta(\omega^2)/\omega_0^2$. We first consider frequencies within the disorder-broadened flat band but outside the central crossover region with the detuning parameter $\eh=(\omega^2-\omega_0^2)/(W\omega_0^2)=0.2$. For simplicity, we label this detuning as $\omega\sim\omega_0$ in the figures. We find that the three localization curves for different $\eta$, plotted in Fig.~\ref{fig:disorder}(a), follow the conventional $\xi\propto W^{-2}$ law for $W\ll W_{\rm band}$~\cite{Anderson1958,MacKinnon1981}. Once disorder exceeds the residual bandwidth, the length approaches a plateau on the scale of $\ellqm$.

In the case of $\eta=0$ (where the clean chain has an exactly flat band), the plateau persists over a broad range of disorder strengths [Fig.~\ref{fig:disorder}(b)], without the conventional weak-disorder scaling law. At $\delta=1$, the plateau even extends to $W=1.9$, where $\xi/\ellqm\simeq6.35$ despite mass variations of up to $95\%$. At each value of $\delta$, changing $W$ leaves the normalized plateau height nearly unchanged.

At the detuned frequency $\omega\sim\omega_0$ (i.e., $\eh=0.2$), the values of $\delta$ in Fig.~\ref{fig:disorder}(c) give $\xi\simeq6.3\ellqm$ at $W=0.002$. The additional points at $W=10^{-4}$ and $0.02$ follow the same trend. The ratio varies by about $4\%$ over $10^{-3}\le\delta\le1$, and a similar collapse is found over $0.2\le|\eh|\le0.3$ for $\delta\le0.3$. In this frequency range, $\xi$ remains on a plateau as $W$ varies, while the height of that plateau changes in proportion to $\ellqm$ as the gap-opening parameter $\delta$ is varied. This is the geometric scaling predicted for quantum metric localization~\cite{Chau2026,Dai2026}. The persistence of these plateaus over a broad range of disorder strengths is also consistent with quantum metric protection~\cite{Dai2026}. The coefficient $6.3$ depends on the detuning and disorder ensemble; it is the proportionality to $\ellqm$ that agrees with the predicted geometric scaling.

\begin{figure}[!t]
\centering
\includegraphics[width=\columnwidth]{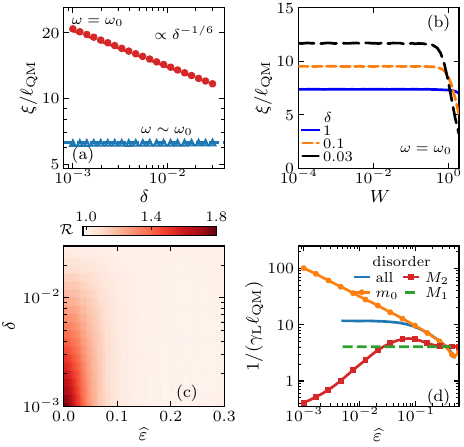}
\caption{\label{fig:omega0}Exceptions to quantum metric scaling close to the clean flat-band frequency. (a) $\xi/\ellqm$ at $W=0.002$ for four independent realizations, comparing $\omega=\omega_0$ with $\omega\sim\omega_0$ ($\eh=0.2$). The constant and $\delta^{-1/6}$ guides correspond to $\xi\propto\delta^{-1/2}$ and $\xi\propto\delta^{-2/3}$, respectively. (b) Localization-length plateaus at $\omega=\omega_0$ versus $W$ for $\delta=1$, $0.1$, and $0.03$: the lengths remain nearly independent of disorder strength while their normalized heights differ. (c) Raw $21\times61$ map at $W=0.002$ of the collapse ratio $\mathcal R$ defined in the text. Scaling with $\ellqm$ gives $\mathcal R=1$; the additional dependence on $\delta$ identifies a central crossover region near $\eh=0$. (d) Inverse transfer rate $1/(\gamma_{\rm L}\ellqm)$ for sublattice-selective disorder at $\delta=0.03$, $W=0.02$. It equals $\xi/\ellqm$ for localized states; the dashed $M_1$ branch describes evanescence.}
\end{figure}

\textit{Exception at the exact flat-band frequency.---} Sufficiently close to $\omega_0$, the localization length still forms a plateau as disorder strength varies, but its dependence on the gap-opening parameter no longer follows the quantum metric length. At the exact flat-band frequency $\omega=\omega_0$, the results from the exact transfer equation for continuous mass disorder give
\begin{equation}
\xi(\omega_0)\propto a\delta^{-2/3},\qquad
\frac{\xi(\omega_0)}{\ellqm}\propto\delta^{-1/6}
\quad(\delta\ll1).
\label{eq:omega0scaling}
\end{equation}
The ratio $\xi/\ellqm$ is no longer a constant and now grows as the gap-opening parameter decreases [Fig.~\ref{fig:omega0}(a)]. Four independent realizations at $W=0.002$ give exponents $0.67$ at $\omega=\omega_0$ and $0.50$ at $\omega\sim\omega_0$ ($\eh=0.2$) over the range $10^{-3}\le\delta\le0.03$.

At the exact flat-band frequency $\omega=\omega_0$, the localization length nevertheless remains nearly constant as $W$ varies [Fig.~\ref{fig:omega0}(b)]. The exception therefore lies in the scaling exponent: the plateau persists as disorder strength varies, but its height can no longer be related to $\ellqm$ by a constant of proportionality. Because $\xi$ remains above $\ellqm$, this anomaly is consistent with quantum metric protection interpreted as a minimum localization scale. 

The change of exponent comes from the way mass disorder couples overlapping ``compact eigenstates'' of the highly degenerate exact flat band. A ``compact eigenstate'' here is an eigenmode whose displacement vanishes outside a finite cluster of sites~\cite{Maimaiti2017}. For the present chain, we denote this unnormalized compact eigenstate as $|n\rangle$, which represents a displacement $-\delta$ on $M_{2,n}$, unit displacement on the resonators in cells $n$ and $n+1$, and zero displacement elsewhere. At the flat-band frequency, the forces on the adjacent $M_1$ masses cancel, so this three-site pattern is an exact eigenmode at $\omega_0$. The index $n$ labels the position of the pattern along the chain: $|n+1\rangle$ is obtained by shifting $|n\rangle$ by one unit cell. We expand the displacement field projected onto the clean flat-band subspace as $u_{\rm FB}=\sum_n c_n|n\rangle$, where $c_n$ specifies the contribution of the compact pattern associated with cell $n$. This gives $u_{2,n}=-\delta c_n$ and $u_{0,n}=s_n=c_{n-1}+c_n$, with $u_{1,n}=0$. The sum in $s_n$ arises because the resonator in cell $n$ is shared by $|n-1\rangle$ and $|n\rangle$. Resonator disorder therefore couples neighboring compact-state amplitudes, whereas $M_2$ disorder contributes a term proportional to $c_n$ in the $n$th projected equation. Projecting the mass-disordered equations of motion onto these states to first order in $W$ gives
\begin{equation}
\begin{aligned}
(\wh_{0,n}+\eh)s_n+(\wh_{0,n+1}+\eh)s_{n+1}\\
+2\delta(\wh_{2,n}+\eh)c_n=0.
\end{aligned}
\label{eq:reduced}
\end{equation}
The overall disorder amplitude cancels from this equation, which explains why a plateau forms but does not require its height to follow the quantum metric. Reducing the disorder strength narrows the frequency splittings without undoing the mixing of flat-band states within the degenerate subspace.

To extract the dependence on $\delta$, put $c_n=(-1)^nd_n$ and define $A_n=\wh_{0,n}+\eh$, $B_n=2\delta(\wh_{2,n}+\eh)$, and $g_n=A_n(d_n-d_{n-1})$. The recurrence becomes $d_n=d_{n-1}+g_n/A_n$ and $g_{n+1}=g_n+B_nd_n$.
For continuous uniform disorder and $|\eh|<1/2$, the probability density of $A_n$ remains finite and nonzero at $A_n=0$. The inverse $1/A_n$ consequently has a Cauchy tail: rare, nearly resonant cells give large increments of $d_n$. While $g$ changes slowly, the typical accumulated displacement scales as $|d_n|\sim n|g|$.

For fixed nonzero $\eh$ as $\delta\to0$, the mean $\langle B_n\rangle=2\delta\eh$ gives $\Delta g\sim\delta\eh n^2g$. The localization length is set by the distance over which this change becomes comparable to $g$, giving $\xi/a\propto(\delta|\eh|)^{-1/2}$. At zero detuning, $\langle B_n\rangle=0$, so the independent $M_2$ fluctuations accumulate instead as $\Delta g\sim\delta n^{3/2}g$, giving $\xi/a\propto\delta^{-2/3}$. Thus the vanishing mean at the flat-band frequency changes the scaling law without restoring any dependence on disorder strength. These scaling balances determine the exponents, not the prefactors.

Independent fluctuations on different sublattices are not required for the anomalous scaling. When all masses within each cell share one uniformly distributed fractional fluctuation, with different cells uncorrelated, the $2/3$ law persists~\cite{SM}. The two ensembles have the same mass-fluctuation distribution on each sublattice and differ only in their within-cell correlations. Those correlations change the prefactors and finite-$\delta$ corrections while preserving the anomalous small-$\delta$ scaling in the tested range. By contrast, binary disorder removes the resonant tails of the inverse coefficients at $\eh=0$ and gives $\xi\propto\delta^{-1}$~\cite{SM}. For continuous uniform disorder, nearly resonant cells and the vanishing mean of the projected diagonal term produce longer localization tails at $\omega_0$. When disorder-induced mixing with dispersive bands is perturbative, the two scaling laws give $\xi(0)/\xi(\eh)\propto\delta^{-1/6}$ at fixed nonzero $\eh$ inside the disorder support.

The balance between the mean and fluctuating contributions also determines how the frequency window around $\omega_0$ narrows. To compare the scaling laws at each frequency, we define $\mathcal R(\delta,\eh)$ as $\xi(\delta,\eh)/\ellqm(\delta)$ divided by its value at $\delta_{\rm ref}=0.03$.
A length proportional to $\ellqm$ gives $\mathcal R=1$, even when the proportionality factor depends on frequency. The ratio shown in Fig.~\ref{fig:omega0}(c) approaches unity outside the central crossover region but reaches $1.79$ at $\eh=0$, $\delta=10^{-3}$. The enhanced length is therefore concentrated close to the clean flat-band frequency. Matching the two rates gives an inner crossover scale $|\eh|\sim\delta^{1/3}$, or $|\eps|\sim W\delta^{1/3}$, so this window narrows as the gap-opening parameter decreases. Quantum metric localization is thus recovered near the flat band, but not within this central crossover region.

\textit{Opposite responses to selective mass disorder.---} Further exceptions to quantum metric scaling arise when disorder is applied selectively to one sublattice. Changing the disordered sublattice leaves the clean quantum metric fixed, but can lead to opposite changes in the localization length [Fig.~\ref{fig:omega0}(d)]. Disorder on $M_1$ leaves the entire flat-band subspace unchanged because every flat-band mode has zero displacement there. For this disorder, the transfer branch away from $\omega_0$ describes evanescence.

Resonator-only disorder preserves the alternating zone-boundary bulk mode: the resonators remain stationary and cannot scatter this displacement pattern. The exact transfer product gives $\gamma_{\rm L}(\omega_0)=0$. For independent uniform resonator disorder, two-sided bounds show that $\xi/a$ diverges as $(\delta|\eh|)^{-1/2}$ up to bounded positive prefactors as $\eh\to0$~\cite{SM}. This result holds at fixed $\delta>0$ and $0<W<2$ without expanding in disorder strength.

Disorder on $M_2$ acts where the same mode has nonzero amplitude and produces the opposite limit. The effective coupling between neighboring $M_2$ masses vanishes at $\omega_0$, while their random local shifts remain finite. For independent uniform $M_2$ disorder, exact bounds give $a\gamma_{\rm L}=\ln(\delta/|\eh|)-1+o(1)$ as $\eh\to0$ at fixed $\delta>0$ and $0<W<2$~\cite{SM}. The localization length becomes arbitrarily short even though states remain present near $\omega_0$; a generic finite realization has no eigenmode exactly at that preassigned frequency. This result concerns the exponential tails; the mode core retains a finite spatial extent. A positive lower bound on the tail length set by $\ellqm$ must therefore specify the allowed disorder operators. When both the resonators and the $M_2$ masses are disordered, the localization length is finite and follows the $\delta^{-2/3}$ law. The same clean geometry supports all three behaviors because disorder acts on different parts of the mode.

In conclusion, we find quantum metric scaling of the localization length and exceptions to this scaling in a classical flat band. Mass disorder produces a localization length proportional to the quantum metric length near, but not too close to, the clean flat-band frequency. Within a narrower central window, resonant coupling between flat-band modes changes the scaling with the gap-opening parameter, even though the localization length remains nearly independent of disorder strength. The enhanced localization length remains consistent with a minimum scale set by the quantum metric. This exception also survives when all masses within each cell share the same fractional fluctuation, with different cells uncorrelated. Selective mass disorder produces further exceptions, allowing the localization length to diverge or the localized tails to become arbitrarily short near the flat-band frequency without changing the clean metric. Quantum metric scaling therefore applies within a specific regime, and its exceptions reflect how disorder couples the flat-band modes.

\begin{acknowledgments}
The authors thank Z.~Q. Zhang and Yixin Xiao for useful discussions. This work
was supported by the Hong Kong Research Grants Council through the Area of Excellence Scheme
(Grant No.~AoE/P-502/20).
\end{acknowledgments}

\textit{Data availability.---} There are no publicly available research data or software supporting this manuscript. Requests for further information or data should be sent to the authors.

\end{document}